\documentclass[10pt,a4paper]{article}
\usepackage{mathrsfs}
\usepackage{epsfig, graphicx}
\usepackage{latexsym,amsfonts,amsbsy,amssymb}
\usepackage{amsmath,amsthm}
\usepackage{color,enumitem}
\usepackage[colorlinks, citecolor=blue]{hyperref}
\usepackage{bm,upgreek}
\usepackage{diagbox}[2011/11/22]
\usepackage{tabularx}
\usepackage{algorithm,algorithmic}
\usepackage{epstopdf} 
\makeatletter % '@' is now a normal "letter" for TeX

\@addtoreset{equation}{section}
\makeatother % '@' is restored as a "non-letter" character for TeX
\allowdisplaybreaks % For long formula

\def\w\eta{\widetilde{\eta}}

\newcommand{\bc}{\begin{center}}
\newcommand{\ec}{\end{center}}
\newcommand{\be}{\begin{eqnarray}}
\newcommand{\ee}{\end{eqnarray}}

\newcommand{\ben}{\begin{eqnarray*}}
\newcommand{\een}{\end{eqnarray*}}

\newtheorem{remark}{Remark}[section]

\begin{document}
\title{Effective approximations for Hartree--Fock exchange potential\footnote{This work was supported by the
Project of Cultivation for young top-motch Talents of Beijing Municipal Institutions (Grant No. BPHR202203022),
the Beijing Municipal Natural Science Foundation (Grant No. 1232001),
and the National Natural Science Foundation of China (Grant No. 12371386). }}
%\title{Convergence acceleration of iteration for nonlinear equations\footnote{This work was supported by the General projects of science and technology plan of Beijing Municipal Education Commission (Grant No. KM202110005011), and the National Natural Science Foundation of China (Grant No. 11801021, 11901047). }}
\author{
Fei Xu\footnote{School of Mathematics, Statistics and Mechanics, Beijing University of Technology, Beijing, 100124, P.R. China.
    (fxu@bjut.edu.cn)}  }

\date{} \maketitle

\begin{abstract}
The Hartree--Fock exchange potential is fundamental for capturing quantum mechanical exchange effects but faces critical challenges in large--scale applications due to its nonlocal and computationally intensive nature.
This study introduces a generalized framework for constructing approximate Fock exchange operators in Hartree--Fock theory, addressing the computational bottlenecks caused by the nonlocal nature. By employing low--rank decomposition and incorporating adjustable variables, the proposed method ensures high accuracy for occupied orbitals while maintaining Hermiticity and structural consistency with the exact Fock exchange operator. Meanwhile, a two--level nested self--consistent field iteration strategy is developed to decouple the exchange operator stabilization (outer loop) and electron density refinement (inner loop), significantly reducing computational costs. Numerical experiments on several molecules demonstrate that the approximate exchange operators achieve near--identical energies compared to that of the exact exchange operator and the NWChem references, with substantial improvements in computational efficiency.


\end{abstract}

\section{Introduction}
The Hartree--Fock (HF) exchange potential \cite{riginal1,original2} occupies a pivotal position in modern computational quantum chemistry, serving as both a cornerstone of wavefunction--based methodologies and a critical component in advanced density functional approximations. As the exact exchange contribution in HF theory establishes the foundation for all post--Hartree--Fock correlation treatments, its inherent nonlocal characteristic encodes essential quantum mechanical effects that govern the electronic structure in both molecules and condensed matter systems. In addition, the integration of the Fock exchange potential into Kohn--Sham density functional theory (KS--DFT) \cite{ks,gga} through hybrid functionals has emerged as a transformative approach, synergistically combining the computational efficiency of DFT with the improved description of electronic exchange effects inherent to wavefunction methods \cite{hybrid,hfintro,hybid3}. This hybridization strategy has demonstrated remarkable success in bridging the theoretical gap between mean--field approximations and exact exchange--correlation treatments, particularly for properties such as band gaps and reaction barriers,
 for which conventional functionals often exhibit systematic errors.

The nonlocal nature of the Fock exchange operator, while theoretically indispensable for capturing quantum mechanical exchange effects, introduces significant computational challenges in practical implementation.
As a self--consistent nonlinear equation rooted in the Hartree--Fock theory, the equation exhibits intrinsic mathematical complexity due to its coupled dependencies. The Fock exchange potentials reciprocally depend on the density matrix, which introduces significant challenges for convergence in iterative solvers -- particularly for metallic systems or those with narrow band gaps. Beyond the fundamental nonlinearity,
the nonlocal character of the operator necessitates manipulation of the density matrix during discretization, generating dense matrices that demand exponential computational complexity for explicit construction. For three--dimensional molecular systems, these challenges collectively result in prohibitive memory and computational costs at each SCF iteration, ultimately limiting the practical applicability of HF--based methodologies.
Overcoming these limitations through efficient numerical strategies that maintain exchange accuracy while reducing operational burdens remains a critical frontier in electronic structure theory.

To tackle these computational bottlenecks, mainstream approaches have diversified, with technical pathways depending on system characteristics and computational objectives.
The classic local Gaussian basis set method uses the Roothaan equation for SCF iterations \cite{guass1,guass2,guass3,guass4,guass5,guassa}. Its key advantage is controllable basis function count: the Hamiltonian matrix can be explicitly constructed and diagonalized directly, with nonlocal exchange integrals supporting analytical computation or pre--storage. However, for non--uniform systems with complex boundaries, its spatial discretization risks numerical instability. Also, fixed basis functions often reduce accuracy despite ease of computation.
Another commonly used method is the plane wave approach \cite{plane3,other2,other1,plane8}. It can efficiently handle periodic systems with the aid of the fast Fourier transform. Plane wave basis are independent of the positions of atomic nuclei, allowing for flexible adjustment of convergence accuracy through truncation strategies. However, it lacks flexibility when dealing with nonperiodic boundaries and complex regions.
Real--space discretization methods (e.g., finite difference, wavelet basis, finite element methods) have become preferred for large--scale simulations \cite{fem1,pask,qui,sun2}. They efficiently handle large--scale systems while maintaining accuracy via adaptive grids and flexible boundaries, converting continuous problems to discrete equations for electronic structure calculations. Yet all these traditional methods face a common bottleneck: direct calculation of nonlocal exchange operator leads to cubic growth in complexity with system size, especially in metallic systems.
To address this, linear scaling methods based on electronic localization offer a new path \cite{ls1,ls4,ls3,ls2}.
They leverage a trait of insulators with a clear HOMO-LUMO gap: occupied orbitals form a compact subspace. A unitary transformation can localize these orbitals in real space; once found, the exchange operator simplifies, enabling efficient large--scale computations. Recent studies \cite{ls5,gap1} show these methods drastically reduce exchange--term costs in large--scale systems with significant band gaps.

Beyond numerical discretization strategies, an alternative approach involves approximating the Fock exchange operator through various techniques.
These include the quantized tensor train
(QTT) representation of the exchange operator \cite{qtt}, Gaussian orbital--based methods, such as the occ--RI--K
method \cite{rik}, the adaptively compressed operator \cite{plane4,plane6}, and approximation via projection operators \cite{plane1,plane2}, etc.
The central insight of these studies is that, for calculating physical properties in HF computations,
it suffices to develop an approximate operator that produces identical results to the exact exchange operator when acting solely on the occupied orbitals.
By focusing computational effort on this subspace, these strategies dynamically reduce the complexity of Fock exchange potential evaluation during iterative solvers while maintaining exactness within the occupied orbital manifold. Critically, this approach has demonstrated potential to overcome historical limitations in large--scale HF simulations by preserving essential quantum mechanical interactions with reduced operational burden.

In this study, we propose a generalized framework for constructing approximate exchange operators that replicate the effect of exact exchange operator on occupied orbitals. Our approach introduces a tunable variable matrix, enabling the creation of approximate exchange operators for arbitrary inputs. By adjusting parameters, established methods such as adaptively compressed exchange and projection--based techniques emerge as special cases within this unified framework. This design ensures exceptional versatility, allowing generation of diverse approximate operators tailored to specific computational requirements. Notably, the performance of the proposed algorithm remains independent of band gap magnitude. This characteristic enables the methodology to be broadly applicable across diverse material systems, spanning from insulators to semiconductors
and even metallic systems. %enabling broad applicability across material systems - from insulators and semiconductors to metallic systems.
In addition, the approximate exchange operator offers significant memory advantages by leveraging low--rank decomposition of the exchange operator. This approach drastically reduces memory footprint in scenarios with high memory demands, particularly for large basis set method.

Once the approximate exchange operator is constructed, subsequent computations become computationally trivial. Consequently, efficient realization of these approximate operators becomes paramount for overall performance. To achieve this, we implement a two--level nested SCF iteration strategy: the outer--loop focuses on optimizing the Fock exchange operator (which contributes relatively little to total energy), requiring only a few iterations, while the inner--loop converges electron density with the exchange operator fixed. This hierarchical approach confines approximate exchange operator construction exclusively to the outer--loop, and the reduced number of outer--loop iterations significantly enhances computational efficiency.

The rest of this paper is organized as follows: In Section 2, we introduce the preliminary knowledge related to the HF method, followed by the approximate Fock exchange operator and the two--level nested SCF iteration. Section 3 presents some numerical experiments to demonstrate the efficiency of the proposed approximate Fock exchange operators. Finally, Section 4 provides some concluding remarks.

\section{Hartree--Fock exchange potential}

\subsection{Self--consistent field iteration for Hartree--Fock equation}
For simplicity, the subsequent discussion in this study focuses on the following HF equation, and extending the analysis to KSDFT with hybrid functionals can be readily achieved:
%For a molecular system comprising $N$ electrons and $M$ nuclei with charges ${Z_1,\cdots,Z_M}$ and locations ${R_1, \cdots, R_M}$, the corresponding HF equation can
%be described by:
\begin{eqnarray}\label{defHartree--Focke}
H[\rho({\bf r}),P({\bf r},{\bf r'})]\phi_\ell:=\big(-\frac{1}{2}\Delta  +V_{\rm ext}
 +V_{\rm Har}[\rho]({\bf r}) +V_{\rm x}[P]({\bf r},{\bf r'})\big)\phi_\ell=\lambda_\ell \phi_\ell,\quad \ell=1,\cdots,N_e,
\end{eqnarray}
where $N_e$ denotes the number of electrons, the density function $\rho({\bf r}):=\sum_{\ell=1}^{N_e}|\phi_\ell({\bf r})|^2$ and
density matrix $P({\bf r},{\bf r'}):=\sum_{\ell=1}^{N_e} \phi_\ell({\bf r})\phi_\ell^*({\bf r'})$, the local potential $V_{\rm ext}$ is used to describe the electron--ion interaction, the Hartree potential
\begin{eqnarray}
V_{\rm Har}[\rho]({\bf r})= \int\frac{\rho({\bf r'})}{|{\bf r}-{\bf r'}|}d{\bf r'}
\end{eqnarray}
and the Fock potential
\begin{eqnarray}\label{vxdef}
(V_{\rm x}[P]\phi_\ell)({\bf r})= -\int\frac{P({\bf r},{\bf r'})}{|{\bf r}-{\bf r'}|}\phi_\ell({\bf r'})d{\bf r'}.% are defined by:
\end{eqnarray}

One common approach to solving the HF equations is to expand the orbitals \(\{\phi_\ell\}_{\ell = 1}^{N_e}\) using a basis set \(\{\psi_{i}\}_{i = 1}^{N}\),
such as Gaussian--type orbitals and numerical atomic orbitals. This leads to a Hamiltonian matrix with a dimension \(N\).

To compute element of the Hamiltonian matrix, the four--center integral
\begin{eqnarray}
\iint\frac{\psi_{i}(\mathbf{r})\psi_{j}^*(\mathbf{r})\psi_{m}(\mathbf{r}')\psi_{n}^*(\mathbf{r}')}{|\mathbf{r}-\mathbf{r}'|}d\mathbf{r}d\mathbf{r}',
\quad i,j,m,n = 1,\ldots,N,
\end{eqnarray}
must be carried out. The cost of the four--center integral is \(O(N^4)\). The quartic scaling becomes very expensive for systems of large sizes.
Therefore, it is crucial to design an efficient algorithm for solving the Fock exchange operator.

Since the HF equation (\ref{defHartree--Focke}) is a nonlinear eigenvalue problem, its solution necessitates employing a SCF iteration method.
This equation needs to be solved in a self--consistent manner until the calculated orbitals match the ones used as input for the Hamiltonian operator.
More precisely, after deriving
the approximate orbits $\phi_\ell^{\rm{old}}$, we first compute the density function $\rho^{\rm{old}}$ and density matrix $P^{\rm{old}}$.
The following linear eigenvalue problem is then solved at each iteration step to obtain the new orbitals until convergence:
\begin{eqnarray}\label{linearized}
\big(-\frac{1}{2}\Delta  +V_{\rm ext}
 +V_{\rm Har}[\rho^{\rm{old}}]({\bf r}) +V_{\rm x}[P^{\rm{old}}]({\bf r},{\bf r'})\big)\phi_\ell=\lambda_\ell \phi_\ell,\quad \ell=1,\cdots,N_e.
\end{eqnarray}

\begin{remark}\label{1stremark}
In this paper, our discussion centers around the Hartree--Fock theory. In the context of KSDFT calculations utilizing hybrid functionals,
for instance, the PBE0 functional, the exchange--correlation energy can be expressed as:
\begin{eqnarray}
E_{\rm{xc}}^{\rm{PBE0}}=\frac{1}{4}E_{\rm x}^{\rm{HF}}+\frac{3}{4}E_{\rm x}^{\rm{PBE}}+E_{\rm c}^{\rm{PBE}}.
\end{eqnarray}
In this equation, \(E_{\rm x}^{\rm{PBE}}\) and \(E_{\rm c}^{\rm{PBE}}\) denote the exchange and correlation parts of the energy originating
from the GGA--type Perdew--Burke--Ernzerhof (PBE) functional \cite{pbe}. As a result, the associated exchange operator is
one--fourth of the Fock exchange operator (\ref{vxdef}).
All methodologies presented in this work for HF equation are directly applicable to hybrid DFT calculations.
\end{remark}
%For the exchange--correlation functionals with screened exchange interactions, like the HSE functional, the exchange-correlation energy is formulated as:
%\begin{eqnarray}
%E_{\rm{xc}}^{\rm{HSE}}(\mu)=\frac{1}{4}E_{\rm x}^{\rm{SR}}(\mu)+\frac{3}{4}E_{\rm x}^{\rm{PBE,SR}}(\mu)+E_{\rm x}^{\rm{PBE,LR}}+E_{\rm c}^{\rm{PBE}}.
%\end{eqnarray}
%In this case, \(E_{X}^{\mathrm{PBE,SR}}\) and \(E_{X}^{\mathrm{PBE,LR}}\) represent the short - range and long - range portions of the exchange contribution within the PBE functional. \(E_{X}^{\mathrm{SR}}\) is the short - range segment of the Fock exchange energy, which is defined as:
%\[E_{X}^{\mathrm{SR}}(\mu)=-\frac{1}{2}\sum_{i,j = 1}^{N_{e}}\iint\psi_{i}(\mathbf{r})\psi_{j}(\mathbf{r})\psi_{j}(\mathbf{r}')\psi_{i}(\mathbf{r}')\frac{\mathrm{erfc}(\mu(|\mathbf{r}-\mathbf{r}'|))}{|\mathbf{r}-\mathbf{r}'|}\mathrm{d}\mathbf{r}\mathrm{d}\mathbf{r}'\] (9)
%Here, \(\mathrm{erfc}\) stands for the complementary error function, and \(\mu\) is a parameter that can be adjusted to regulate the screening length of the short - range part of the Fock exchange interaction. The sole change involves substituting the Coulomb kernel with the screened Coulomb kernel, and the screened Coulomb kernel should be used to define the exchange operator \(V_{X}^{\mathrm{HSE}}\).

%\section{Approximate Fock exchange operator}
\subsection{Approximate Fock exchange operator}
To enhance the computational efficiency of the HF equation, it is essential to address the processing of the Fock exchange operator. In order to reduce the computational costs, we propose a novel class of approximate Fock exchange operators that are more straightforward to solve.

Since the exchange operator is a full--rank operator, the direct application of compression techniques such as singular value decomposition may lead to inaccurate results. Notably, these techniques seek to identify approximate operators that minimize approximation errors for any orbitals. The core concept of our approach is to construct an approximate operator that maintains minimal approximation errors specifically for occupied orbitals. This orbital--adaptive design ensures that upon achieving self--consistency, all physical quantities reproduce those from exact HF calculations.

%Our objective is to construct an approximate operator $\widetilde{V}_{\rm x}$ which satisfies:\\ (1) Hermitian operator:
%\begin{eqnarray}\label{acecond}
%\widetilde{V}_{\rm x} (\mathbf{r}, \mathbf{r}^{\prime} )=\widetilde{V}_{\rm x}^* (\mathbf{r}^{\prime}, \mathbf{r} ).
% \end{eqnarray}
%(2) The results acting on the  occupied orbitals $\{\phi_\ell\}_{\ell=1}^{N_e}$ are consistent with the structures of the exact Fock exchange operator:
%\begin{eqnarray}\label{acecond2}
%(\widetilde{V}_{\rm x} \phi_\ell )(\mathbf{r})=(V_{\rm x} \phi_\ell )(\mathbf{r}).
% \end{eqnarray}

{
Our objective is to construct an approximate exchange operator $\widetilde{V}_{\rm x}$ that satisfies two key properties:

\begin{enumerate}
\item \textbf{Hermitian Property}: The operator maintains the Hermitian symmetry required for physical operators:
\begin{equation}
\widetilde{V}_{\rm x}(\mathbf{r}, \mathbf{r}') = \widetilde{V}_{\rm x}^*(\mathbf{r}', \mathbf{r}).
\label{acecond}
\end{equation}

\vspace{12pt}

\item \textbf{Occupied Orbital Consistency}: For all occupied orbitals $\{\phi_\ell\}_{\ell=1}^{N_e}$, the approximate operator reproduces the exact Fock exchange operator results:
\begin{equation}
(\widetilde{V}_{\rm x} \phi_\ell)(\mathbf{r}) = (V_{\rm x} \phi_\ell)(\mathbf{r}).
\label{acecond2}
\end{equation}
\end{enumerate}

}

For this purpose, we first compute function $w_\ell(\mathbf{r})$ for a specified set of approximate occupied orbitals $\{\phi_\ell\}_{\ell=1}^{N_e}$ as follows:
\begin{eqnarray}\label{def4w}
w_\ell(\mathbf{r})= (V_{\rm x} \phi_\ell )(\mathbf{r}), \quad \ell=1, \ldots, N_e.
\end{eqnarray}
Then, we assume the approximate exchange operator satisfies the following form:
\begin{eqnarray}\label{vxdef2}
\widetilde{V}_{\rm x}(\mathbf{r}, \mathbf{r}^{\prime} )=[W(\mathbf{r}),\Phi(\mathbf{r})]A[W(\mathbf{r}'),\Phi(\mathbf{r}')]^*,
\end{eqnarray}
where the function vectors $W$ and $\Phi$ are defined by $W=(w_1,\cdots,w_{N_e})$, $\Phi=(\phi_1,\cdots,\phi_{N_e})$,
$[W,\Phi]$ denotes a vector formed by splicing together $W$ and $\Phi$, and $A$ denotes a $2N_e\times 2N_e$--matrix.

It is worth emphasizing that there exist other types of symmetric forms. However, (\ref{vxdef2}) is a form that balances theoretical generality with algorithmic simplicity.

Given that $[W,\Phi]$ is a known vector, the derivation of matrix $A$ is essential for constructing the approximate operator $\widetilde{V}_{\rm x}$.
To determine the elements of  $A$, we apply the conditions in (\ref{acecond}) and (\ref{acecond2}).

The first condition (\ref{acecond}) imposes that $A$ must be Hermitian, ensuring physical consistency of the operator.
To satisfy the second condition (\ref{acecond2}), we further compute
\begin{eqnarray}
\widetilde{V}_{\rm x}\Phi=[W(\mathbf{r}),\Phi(\mathbf{r})]A   \begin{bmatrix}
(w_1^*(\mathbf{r}'),\phi_1(\mathbf{r}')) & \cdots & (w_1^*(\mathbf{r}'),\phi_{N_e}(\mathbf{r}')) \\
\vdots & \vdots & \vdots \\
(w_{N_e}^*(\mathbf{r}'),\phi_1(\mathbf{r}')) & \cdots & (w_{N_e}^*(\mathbf{r}'),\phi_{N_e}(\mathbf{r}')) \\
1 & \cdots & 0 \\
\vdots & \vdots & \vdots \\
0 & \cdots & 1 \\
\end{bmatrix},
\end{eqnarray}
where the orthognality $(\phi_i^*(\mathbf{r}'),\phi_j(\mathbf{r}')):=\int\phi_i^*(\mathbf{r}')\phi_j(\mathbf{r}')d\mathbf{r}'=\delta_{ij}$ is used.

%\begin{eqnarray}
%\widetilde{V}_{\rm x}\phi_i&=&[W(y),\Phi(y)]A[(w_1(x),\phi_i(x)),\cdots,(w_{N_e}(x),\phi_i(x)),(\phi_1(x),\phi_i(x)),\cdots,(\phi_1(x),\phi_i(x))]^* \nonumber\\
%&=&[W(y),\Phi(y)]A[(w_1(x),\phi_i(x)),\cdots,(w_{N_e}(x),\phi_i(x)),0,\cdots,1,\cdots,0]^*,
%\end{eqnarray}
%As $\widetilde{V}_{\rm x}\phi_i=w_i(y)$, we have

For simplicity, we define
\begin{eqnarray}\label{def4m}
M= \begin{bmatrix}
(w_1^*(\mathbf{r}'),\phi_1(\mathbf{r}')) & \cdots & (w_1^*(\mathbf{r}'),\phi_{N_e}(\mathbf{r}')) \\
\vdots & \vdots & \vdots \\
(w_{N_e}^*(\mathbf{r}'),\phi_1(\mathbf{r}')) & \cdots & (w_{N_e}^*(\mathbf{r}'),\phi_{N_e}(\mathbf{r}')) \\
\end{bmatrix}.
\end{eqnarray}
Then based on the definitions of $w_\ell(\mathbf{r})$ and $V_{\rm x}$, we can find that $M$ is a Hermitian matrix. As the condition (\ref{acecond2}) requires $\widetilde{V}_{\rm x}\Phi=W(\mathbf{r})$, we have
\begin{eqnarray}
[W(\mathbf{r}),\Phi(\mathbf{r})]A \begin{bmatrix}
M  \\
I \\
\end{bmatrix}     =W(\mathbf{r}).
\end{eqnarray}

Next, let's divide $A$ into four blocks

\begin{eqnarray}
A= \begin{bmatrix}
A_{11} & A_{12} \\
A^*_{12} & A_{22} \\
\end{bmatrix},
\end{eqnarray}
where each block is a $N_e\times N_e$--matrix, and $A_{11}$, $A_{22}$ are Hermitian matrices.

Then we have
\begin{eqnarray}
[W(\mathbf{r}),\Phi(\mathbf{r})] \begin{bmatrix}
A_{11} & A_{12} \\
A^*_{12} & A_{22} \\
\end{bmatrix}  \begin{bmatrix}
M  \\
I \\
\end{bmatrix}          =W(\mathbf{r}),
\end{eqnarray}
which further yields:
%\begin{eqnarray}
%W(\mathbf{r})A_{11}M+\Phi(\mathbf{r})A^*_{12}M+W(\mathbf{r})A_{12}+\Phi(\mathbf{r})A_{22} =W(\mathbf{r}),
%\end{eqnarray}
%That is
\begin{eqnarray}
W(\mathbf{r})(A_{11}M+A_{12})+\Phi(\mathbf{r})(A^*_{12}M+A_{22}) =W(\mathbf{r}).
\end{eqnarray}
By equating the coefficients of the function on both sides, we can obtain:
\begin{align}\label{system}
\begin{cases}
A_{11}M+A_{12} &= I, \\
A^*_{12}M+A_{22} &= 0.
\end{cases}
\end{align}
Let us treat $A_{11}$ as a free variable that can be chosen as any Hermitian matrix. % and the equation becomes:
Then from the first equation of (\ref{system}), we can derive
 \begin{eqnarray}
A_{12}=I-A_{11}M.
\end{eqnarray}
Next, from the second equation of (\ref{system}), we can derive
 \begin{eqnarray}
A_{22}=-(I-A_{11}M)^*M=MA_{11}M-M.
\end{eqnarray}

Taking these expressions into (\ref{vxdef2}),
the approximate Fock exchange operator can be expressed by:
\begin{eqnarray}\label{fockapp}
\widetilde{V}_{\rm x}(\mathbf{r},\mathbf{r}') &=& W(\mathbf{r})A_{11}W^*(\mathbf{r}')+W(\mathbf{r})(I-A_{11}M)\Phi^*(\mathbf{r}') \nonumber\\
&&+\Phi(\mathbf{r})(I-MA_{11})W^*(\mathbf{r}')+\Phi(\mathbf{r})(MA_{11}M-M)\Phi^*(\mathbf{r}'),
\end{eqnarray}
where $A_{11}$ is any Hermitian matrix.

\begin{remark}
Since \(A_{11}\) can be any Hermitian matrix, there exist infinitely many choices for matrix $A$, thereby enabling the construction of infinitely many approximate Fock exchange operators. Below, we present several illustrative examples, though the optimal selection of \(A_{11}\) remains an open problem.

\noindent{\bf Case 1: \( A_{11} = 0 \)}\\
The approximate operator takes the form:
\begin{eqnarray}\label{case1}
\widetilde{V}_{\rm x}(\mathbf{r},\mathbf{r}') = W(\mathbf{r})\Phi^*(\mathbf{r}') + \Phi(\mathbf{r})W^*(\mathbf{r}') - \Phi(\mathbf{r})M\Phi^*(\mathbf{r}').
\end{eqnarray}

\noindent{\bf Case 2: \( A_{11} = I \)}\\
The operator becomes:
\begin{eqnarray}\label{case2}
\widetilde{V}_{\rm x}(\mathbf{r},\mathbf{r}') = \ W(\mathbf{r})W^*(\mathbf{r}') + W(\mathbf{r})(I - M)\Phi^*(\mathbf{r}')
+ \Phi(\mathbf{r})(I - M)W^*(\mathbf{r}') + \Phi(\mathbf{r})(M^2 - M)\Phi^*(\mathbf{r}').
\end{eqnarray}

\noindent{\bf Case 3: \( M \) is invertible (\( A_{11} = M^{-1} \))}\\
The operator simplifies to:
\begin{eqnarray}\label{case3}
\widetilde{V}_{\rm x}(\mathbf{r},\mathbf{r}') = W(\mathbf{r})M^{-1}W^*(\mathbf{r}').
\end{eqnarray}

\noindent{\bf Case 4: \( M \) is singular (non--invertible)}\\
For a singular Hermitian matrix \( M \), we use its unitary diagonalization:
\[
M = U^*\Lambda U,
\]
where \( U \) is unitary, and \( \Lambda = \text{diag}(\lambda_1, \dots, \lambda_{N_e}) \) contains the eigenvalues of \( M \). Define the pseudoinverse matrix \( \Lambda^+ \) as:
\[
\Lambda^+ = \text{diag}(\lambda_1^+, \dots, \lambda_{N_e}^+), \quad \text{with} \quad
\lambda_i^+ =
\begin{cases}
\lambda_i^{-1} & \text{if } \lambda_i \neq 0, \\
0 & \text{if } \lambda_i = 0.
\end{cases}
\]
Choosing \( A_{11} = U^*\Lambda^+U \), and defining \( \widetilde{U} = (I - \Lambda\Lambda^+)U \), we obtain:
\[
I - A_{11}M = I - MA_{11} = \widetilde{U}^*\widetilde{U}.
\]
The approximate exchange operator then becomes:
\begin{eqnarray}\label{224}
\widetilde{V}_{\rm x}(\mathbf{r}, \mathbf{r}') = \ W(\mathbf{r})U^*\Lambda^+UW^*(\mathbf{r}')
+ W(\mathbf{r})\widetilde{U}^*\widetilde{U}\Phi^*(\mathbf{r}') + \Phi(\mathbf{r})\widetilde{U}^*\widetilde{U}W^*(\mathbf{r}').
\end{eqnarray}

\end{remark}

\subsection{Two--level nested SCF iteration}
After constructing the approximate exchange operator, subsequent calculations exhibit low computational complexity. However, the per--iteration construction of approximate operators remains computationally intensive due to the need to evaluate $W(\mathbf{r})$ and $M$ via (\ref{def4w}) and (\ref{def4m}) before each SCF iteration. To improve computational efficiency, we focus on reducing the total number of iterations required for convergence.

To address this challenge, we employ a two--level nested iteration strategy that enhances efficiency in discretizing the Fock exchange operator.
This approach strategically decouples the iterative process into two interconnected levels, an outer loop and an inner loop, each tailored to address distinct aspects of the exchange operator’s behavior.

The outer SCF iteration is dedicated to stabilizing the Fock exchange operator, which, despite representing a relatively small fraction of the total energy,
involves computationally intensive many--electron integrals. By fixing the orbitals that define Fock exchange operator during the inner loop iterations, the outer loop minimizes redundant recalculations of the Fock matrix. This freezing of Fock exchange operator allows the Hamiltonian to depend solely on the electron density, which is iteratively refined in the inner loop. Convergence of the outer iteration is monitored through the Fock exchange energy, ensuring that updates to Fock exchange operator occur only after the inner loop has sufficiently converged density.
Since the Fock exchange energy constitutes a relatively small proportion of the total energy, the outer iteration can converge with only a few iterations.

The inner SCF iteration focuses on efficiently updating the electron density under the pre--converged Fock exchange operator.
This decoupling permits the use of advanced charge mixing schemes, such as Anderson acceleration or Pulay’s direct inversion of the iterative subspace (DIIS), to rapidly converge the density without recalculating the exchange contributions. By treating the density updates as a nonlinear eigenvalue problem, the inner loop achieves efficient refinement of wavefunctions and energies.
%Notably, this methodology avoids direct manipulation of the density matrix, which would be computationally prohibitive for large basis sets.
By isolating the complexity of the Fock exchange operator to the outer loop, this approach provides a robust solution for accelerating HF self--consistency in systems where exchange effects are critical.

The detailed process of the approximate exchange operator technique coupled with two--level nested SCF iteration is shown in Algorithm \ref{Multilevel CorrectionnestedACE}.
\begin{algorithm}[h]\caption{Approximate exchange operator coupled with two--level nested SCF iteration}
\begin{algorithmic}[1]
%\REQUIRE Initial value $\phi_{\ell,h_{k+1}}^{(1,1)}, \ \ell=1,2,\cdots,N$, tolerance $TOL$.
%\ENSURE $(\lambda_{\ell,h_{k+1}}, \phi_{\ell,h_{k+1}}), \ \ell=1,2,\cdots,N.$
%\STATE Set $p=1, q=1$.
\WHILE{Exchange energy is not converged}
\STATE Compute the vector $W$ accoridng to (\ref{def4w}).
\STATE Assemble the matrix $M$ according to (\ref{def4m}).
\STATE Update the Fock exchange operator $\widetilde{V}_{\rm x}$ according to (\ref{fockapp}).
\WHILE{Density function is not converged}
\STATE Solve the HF equation with the approximate Fock exchange operator $\widetilde{V}_{\rm x}$.
%\STATE Update the density function.
\STATE Update the Hartree matrix.
\ENDWHILE
%\STATE Compute the exchange energy.
\STATE Update the density matrix.
%\STATE Set $q=p$.
\ENDWHILE
\end{algorithmic}
\label{Multilevel CorrectionnestedACE}
\end{algorithm}

\section{Numerical experiments}
This section presents numerical experiments on several molecules to validate the efficiency of the proposed approximate Fock exchange operator.
We exhibit the efficiency from two points. The fists one is accuracy. We compare the numerical results of our algorithm with
that of the exact Fock exchange operator and the reference values from software NWChem.
The second one focus on the computational time.
To ensure a fair and consistent comparison across different approximate exchange operators, we employed the same numerical discretization approach in all experimental configurations.
Specifically, the finite element method with piecewise linear polynomial basis functions was systematically applied for spatial discretization across all test cases.
This methodological consistency allows us to rigorously evaluate the performance differences arising solely from the exchange operator approximations rather than variations in the discretization scheme.

%To be fair, we tested the same numerical method for different approximate exchange operator adopted in our numerical experiments, i.e. the finite
%element method with linear polynomial basis for all the numerical experiments.

\subsection{Precision}
In the first subsection, we focus on validating the accuracy of the proposed approximate Fock exchange operators.
The molecular HLi, C$_2$H$_6$, C$_6$H$_6$ and C$_6$H$_{12}$O$_6$ are adopted to demonstrate the numerical results.
To this end, we report the energies from the approximate exchange operator, exact exchange operators and the values from NWChem. The approximate Fock
exchange operators (\ref{case1}), (\ref{case2}), (\ref{224}) in cases 1, 2, 4 are used. The corresponding numerical results are presented in Tables \ref{timetableex11}--\ref{timetableex13}.
The symbol ``--" means the computer runs out of memory before we reach the desired accuracy.
We can find that the energies calculated using the three types of approximate Fock exchange operators all demonstrate high accuracy when compared to the reference values.
Meanwhile, the energy of the approximate exchange operator in the finite element framework will eventually converge to that of the exact exchange operator in the finite element framework.

Additionally, it is observed that as molecular systems grow in complexity, the computational demands of exact exchange operators in  finite element method exceed the processing capacity of computers, highlighting the superior computational efficiency and memory savings achieved by the approximate operators.

\begin{table}[htbp]
\begin{center}
\resizebox{0.83\textwidth}{!}{
\begin{tabular}{c | c c c c}\hline
Atom & Energy with approximate operator &Energy with exact operator & Energy of NWChem \\ \hline
%Helium  &3.114296E+5    & 1.093824E+2  & 2.947164E+3 \\ %\hline
HLi  & --7.984251 &--7.984258 & --7.984234 \\ %\hline
C$_2$H$_6$          &  --79.237340 &--79.237346& --79.237278\\ %\hline
C$_6$H$_6$   &  --230.729521  &--& --230.728364\\ %\hline
C$_6$H$_{12}$O$_6$         & --683.389147	 &--& --683.379984	\\ \hline
\end{tabular}
}
\end{center}
\caption{The energies with approximate exchange operator in case 1, exact exchange operator and the reference values from NWChem.}
\label{timetableex11}
\end{table}

\begin{table}[htbp]
\begin{center}
\resizebox{0.83\textwidth}{!}{
\begin{tabular}{c | c c c c}\hline
Atom & Energy with approximate operator &Energy with exact operator & Energy of NWChem \\ \hline
%Helium  &3.114296E+5    & 1.093824E+2  & 2.947164E+3 \\ %\hline
HLi  & --7.984253 &--7.984258 & --7.984234 \\ %\hline
C$_2$H$_6$          &  --79.237341 &--79.237346& --79.237278\\ %\hline
C$_6$H$_6$   &  --230.729526  &--& --230.728364\\ %\hline
C$_6$H$_{12}$O$_6$         & --683.389143	 &--& --683.379984	\\ \hline
\end{tabular}
}
\end{center}
\caption{The energies with approximate exchange operator in case 2, exact exchange operator and the reference values from NWChem.}
\label{timetableex12}
\end{table}

\begin{table}[htbp]
\begin{center}
\resizebox{0.83\textwidth}{!}{
\begin{tabular}{c | c c c c}\hline
Atom & Energy with approximate operator &Energy with exact operator & Energy of NWChem \\ \hline
HLi  & --7.984256 &--7.984258 & --7.984234 \\ %\hline
C$_2$H$_6$          &  --79.237349 &--79.237346& --79.237278\\ %\hline
C$_6$H$_6$   &  --230.729524  &--& --230.728364\\ %\hline
C$_6$H$_{12}$O$_6$         & --683.389145	 &--& --683.379984	\\ \hline
\end{tabular}
}
\end{center}
\caption{The energies with approximate exchange operator in case 4, exact exchange operator and the reference values from NWChem.}
\label{timetableex13}
\end{table}

\subsection{Computational efficiency}
\begin{figure}[htb]
\centering
\includegraphics[width=5cm]{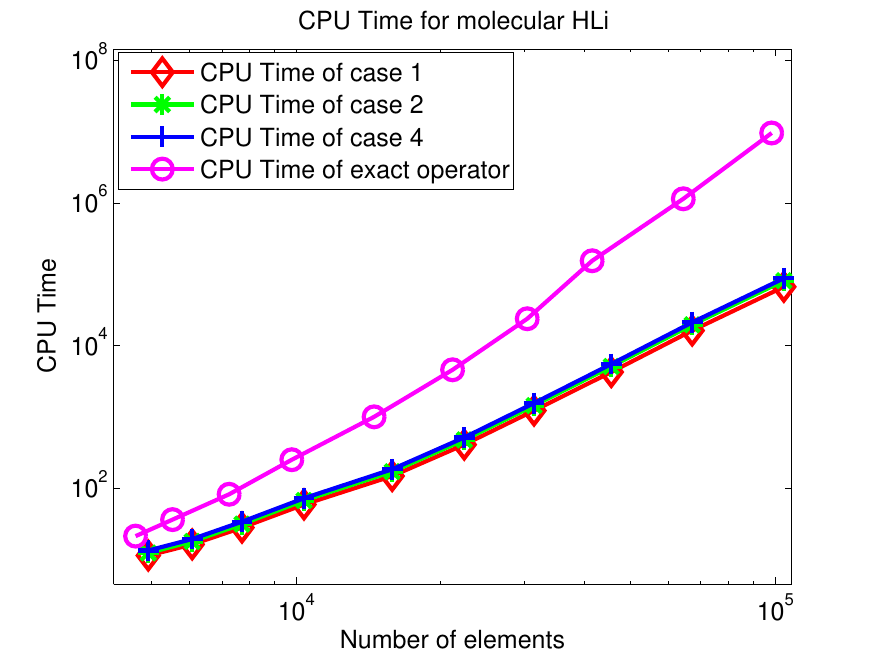} \ \ \ \ \ \ \
\includegraphics[width=5cm]{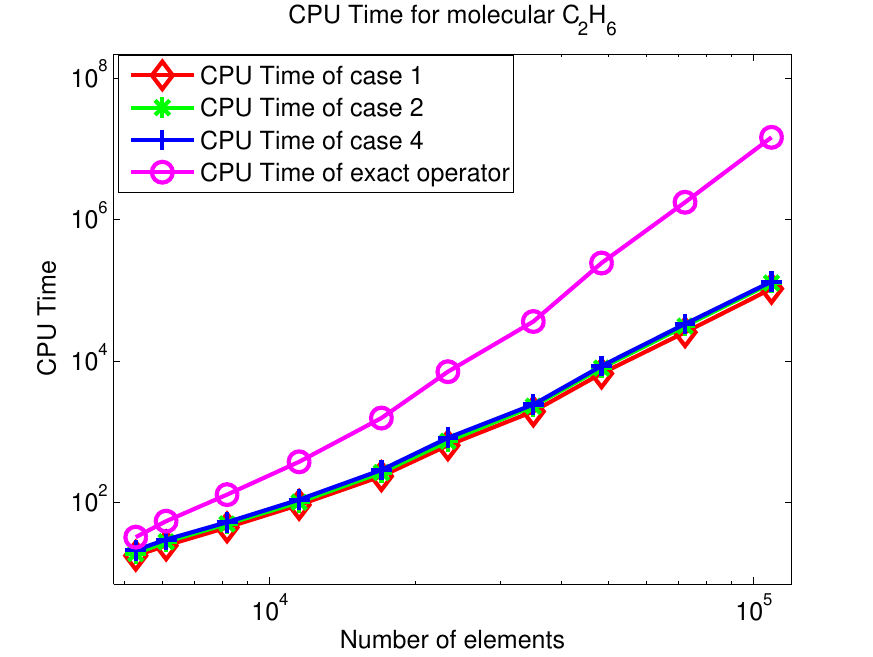}\\
\includegraphics[width=5cm]{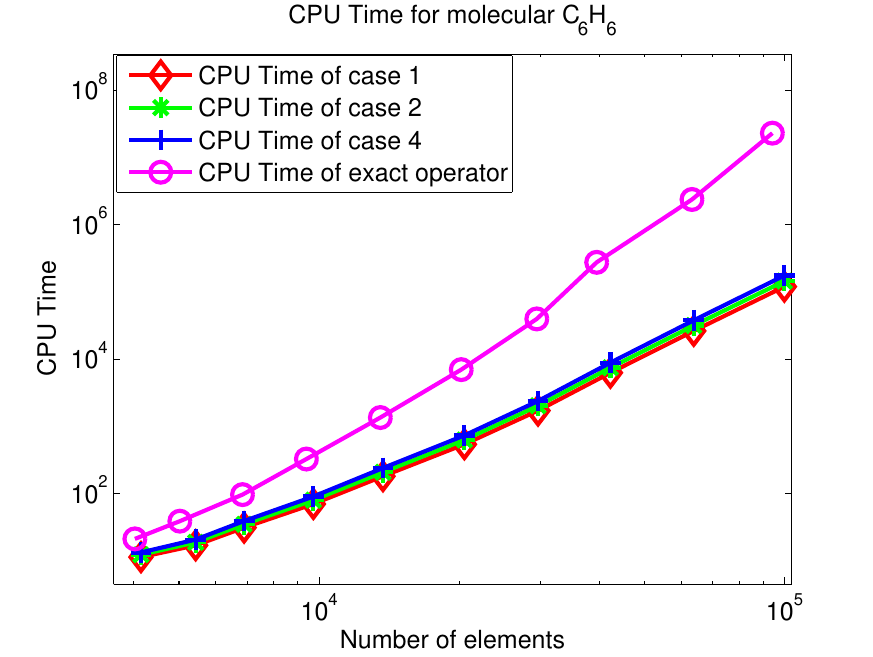}\ \ \ \ \ \ \
\includegraphics[width=5cm]{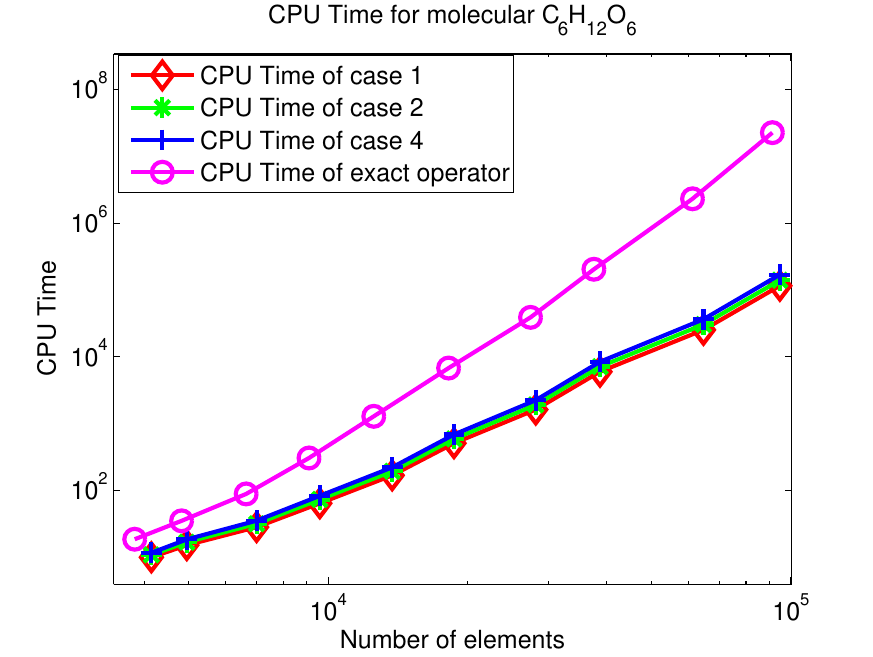}
\caption{The computational time with approximate exchange operators presented in cases 1, 2, 4 and the exact exchange operator.}
\label{timeplot}
\end{figure}

In this subsection, we evaluate the computational efficiency of different approximate Fock exchange operators. To ensure a fair comparison, we conducted tests on these different operators within an identical finite element framework, with the corresponding numerical results presented in Figure \ref{timeplot}.
Figure  \ref{timeplot} illustrates the computational time when using the exact exchange operator and the approximate exchange operators.
It is evident that employing such approximate operators lead to significant reduction in computational time.

%Our algorithm can generate infinitely many types of approximate Fock exchange operators to meet the needs of diverse application scenarios.
%This study only conducts preliminary verification on several basic scenarios, and the results show that the performance differences among the selected operators in conventional problems are minimal.
%It is worth noting that when dealing with complex problems that may arise in advanced applications, specially designed special approximate operators may be required. Compared with general schemes, such operators are expected to achieve significant efficiency improvements.

\section{Conclusion}

The Fock exchange operator constitutes a cornerstone of HF theory, playing a pivotal role in accurately describing quantum mechanical exchange effects essential for predicting electronic structures in molecules and condensed matter systems. However, its nonlocal nature introduces prohibitive computational costs, particularly in large--scale applications.
This work overcomes these limitations by proposing an adjustable, low--rank approximate exchange operator framework that preserves essential quantum interactions while significantly enhancing computational efficiency as the system size increases. The two--level nested SCF iteration strategy enhances computational efficiency by isolating the complexity of the Fock operator to the outer loop, allowing rapid convergence of electron density in the inner loop via advanced charge mixing techniques. Numerical validations confirm that the approximate Fock exchange operators maintain high accuracy for physical properties while drastically reducing computational time. These advancements provide a versatile solution for accelerating HF--based electronic structure calculations in quantum chemistry and materials science.

While the presented scheme demonstrates significant improvements in computational efficiency, a rigorous convergence theory remains elusive.
Regarding the selection of free parameter $A_{11}$, whether a global optimal strategy exists or customized settings based on model--specific characteristics are required remains an open problem lacking systematic conclusions.
Our future work will focus on establishing a convergence proof framework and developing universal selection criteria for parameter $A_{11}$.
%------------------------------------------------------------------------------------------------------

%------------------------------------------------------------------------------------------------------

\end{document}